# STELLAR SOURCES OF GAMMA-BURSTS

B.I. Luchkov
National Research Nuclear University MEPhI
Moscow, Kashirskay Road 31

Our correlation analysis of Swift gamma-ray burst coordinates and nearby star locations (catalog Gliese) reveals 4 coincidences with angular accuracy better than $0.1°$. The random probability is $2×10^{-4}$, so evidencing that coincident stars are indeed gamma-ray burst sources. The fifth coincident gamma-burst could be added (angular accuracy $0.14°$) with random probability for 5 events $\sim 10^{-5}$.

Cosmic gamma-ray bursts (CGB), short duration gamma-ray fluxes with energies from 10 KeV to 10 MeV registered by detectors on satellites and spaceships, were mystery for a long times [1, 2]. The CGB nature was unknown: what are there sources, at what distances and in what processes they are produced? Their resolution improvement to angular minutes came to discovery of source X-ray and optical afterglow. Large ground telescopes and Hubble space telescope could find fading transient objects. Measuring in some cases redshift Z established the transient nature [3]. About 40 % of them have $Z \geq 1$ and consequently are on large (cosmological) distances appearing as fireballs – Supernova flares at remote galaxies [4]. However the gamma-burst mystery was not entirely revealed: what are the other 60 % ?

Coming from idea that the CGB sources could present not single class but summery of different objects according their power, location and emitting processes, an assumption was given about near sources [5] among which could be active (flaring) stars of small masses. They are young stars of late spectral classes [6].

This work is devoted to continuation of search stars which flares could generate gamma-ray bursts.

**Correlation analysis of gamma-ray burst and star locations**

Gamma-burst positions are taken from Swift catalog [7] which contain now (April 2009) 276 unknown CGBs and 128 cosmological CGBs with Z from 0.5 to 5. Swift angular accuracy is $\sigma = 0.1°$. CGB coordinates are compared with coordinates of nearby stars from Gliese catalog [8]. In order to decrease number of random applications the stars are taken with parallaxes $P > 0.05$ what correspond to distances $R \leq 20$ pc and only of G, K, M spectral classes which eventually could be gamma-ray burst sources due to their flare activity. Thus number of stars taken for correlation with gamma-bursts was $N_{st} = 1340$. The cosmological CGBs which could coincide only randomly serve as a comparison criterion.

As numerical indicator for CGB and star coordinates coincidence a deviation $\Delta r = (\Delta\alpha^2 + \Delta\delta^2)^{1/2}$ was used, where $\Delta\alpha$ and $\Delta\delta$ are angular differences for right ascension and declination. The analysis result is given in table 1.

Table 1. The most small deviation $\Delta r$ (degree)

|  | 0-0.06 | 0.06-0.12 | 0.12-0.18 | 0.18-0.24 | 0.24-0.30 | 0.30-0.36 | 0.36-0.42 | 0.42-0.48 |
|---|---|---|---|---|---|---|---|---|
| CGB | 1 | 3 | 1 | 2 | 2 | 4 | 2 | 3 |
| Cosmological CGB (Z) | 0 | 0 | 0 | 0 | 2 | 1 | 0 | 2 |

As one can see 4 coincidences of CGB and stars coordinates were found with $\Delta r \leq$

0.1º what correspond to Swift angular resolution. There are no similar near coincidence among control group of cosmological CGBs. Deviations Δr > 0.18º seen both in investigating and control groups are evidently background of random application.

Four CGB (Δr =0.076º, Δr = 0.054º, Δr = 0.081º, Δr = 0.123º) are evidently of stellar origin. The fifth event (Δr = 0.139º) is not so obvious but lack of control events in this interval gives some indication in favor of its stellar nature.

The mean value $\Delta r_m = 0.084º$ for 4 coincident CGBs is in good agreement with Swift angular resolution. Let us obtain the probability of their random application. The number of coincident evens is $N = S_{coin} N_b N_{st} / \Omega = 0.3$, where $S_{coin} = \pi \sigma^2 = 0.03$ degree$^2$ – square of every event, $N_b = 276$, $N_{st} = 1340$, $\Omega = 41253$ degree$^2$ – total sky surface. The Poisson probability is equal $W = e^{-N} N^4 / 4! = 2 \times 10^{-4}$.

The same calculation could be maid for 5 coincident events. In this case N = 0.2 and probability $W = 10^{-5}$. Both values are small enough what gives definite evidence for real stellar CGB identification.

**Found stellar CGB sources**

The list of coincident stars present in table 2.

Table 2. Stars coincident with CGBs

| Star (Gliese) | Spectral class | Parallax P | Coincident accuracy Δr (º) | CGB |
|---|---|---|---|---|
| NN 3779 | M 3.5 | 0.072 | 0.054 | 50522 |
| GJ 1243 | m | 0.083 | 0.081 | 60105 |
| G 1241 | dK6 | 0.057 | 0.076 | 080319D |
| NN 3929 | M 6 | 0.061 | 0.123 | 90404 |
| G 1653 A, B | G 8v | 0.054 | 0.139 | 70309 |

Found stellar sources of gamma-ray bursts are objects of late spectral classes (G, K, M), as have to be for active flaring stars. Their parallaxes (P > 0.05) and distances (R < 20 pc) are caused due to necessity to decrease rate of random applications. Certainly stellar sources must be at greater distances but they could be extracted only by detectors with higher angular resolution. Our obtained data have not given the exact share of stellar CGB sources. It is not excluded that they may constitute another 60 % of CGBs.